\documentclass[12pt]{article}
\usepackage{latexsym}

\def\ofo{ { {}_2 \! F_1 }}

\newcommand{\beq}{\begin{equation}}
\newcommand{\eeq}[1]{\label{#1}\end{equation}}
\newcommand{\bea}{\begin{eqnarray}}
\newcommand{\eea}[1]{\label{#1}\end{eqnarray}}

\begin{document}
\begin{titlepage}
\hfill NYU-TH/02/01/01

\hfill CU-TP-1049

\hfill  hep-th/0201261

\vspace{20pt}

\begin{center}
{\large\bf{ON THE GRAVITON SELF ENERGY IN $ADS_4$}}
\end{center}

\vspace{6pt}

\begin{center}
{\large M. Porrati$^1$ and A. Starinets$^2$} \vspace{20pt}

{\em $^1$ Department of Physics, NYU, 4 Washington Pl, New York NY 10003}\\
\vspace{6pt}
{\em $^2$ Department of Physics, Columbia University, New York NY 10027}
\end{center}

\vspace{12pt}

\begin{center}
\textbf{Abstract}
\end{center}
\begin{quotation}\noindent
We consider 
Einstein gravity coupled to a CFT made of a single {\em free} conformal 
scalar in $AdS_4$. 
This simple case is rich enough to explain an unexpected gravitational
Higgs phenomenon that has no flat-space counterpart, yet simple enough that
many calculations can be carried on exactly. Specifically, in this paper
we compute the graviton self energy due to matter, and we exhibit its spectral 
representation. This enables us to find the spin-2 bound-state content of the
system.
\end{quotation}
\vfill
\end{titlepage} 
\section{Introduction}
Gravity in Anti de Sitter space exhibits several unusual properties. In our 
view, one of the most surprising is that gravity coupled to a 
conformal field theory can become massive. This was seen in strongly 
self-interacting CFTs in~\cite{kr,p2}, and in~\cite{p3} in a {\em free} CFT. 
This phenomenon is peculiar to AdS in two ways.

First of all, a gravitational Higgs phenomenon requires the existence of a
Goldstone boson. In $AdS_4$, the Goldstone boson is a massive vector, 
belonging to the unitary irreducible representation called $D(4,1)$~\cite{p3} 
(see e.g~\cite{e,n} for notations). In flat space, a free theory, as the CFT 
considered here and in~\cite{p3}, cannot form any bound state so it cannot 
supply the Goldstone boson. In $AdS_4$ instead, even a free theory
can form  bound states, owing to the discreteness of the energy spectrum. 
Moreover, as shown explicitly in~\cite{p3}, the $D(4,1)$ is a bound
state of a free CFT only when non-reflecting boundary conditions are imposed
on its elementary fields. Boundary conditions can influence bulk physics, since
in AdS null geodesics take a finite coordinate time to reach the boundary.
Again, this cannot happen in flat space. The mass of the graviton must 
therefore vanish with the cosmological constant $\Lambda$; indeed, it
is $O(\sqrt{G}\Lambda)$, where $G$ is the Newton constant~\cite{kr,p2,p3}. 

In Ref.~\cite{p3}, a convenient representation for the matter contribution to
the graviton self energy was given, in the case that 
matter is a free conformal scalar.
That representation uses the techniques of Ref.~\cite{f}. In this paper, we
will review the derivation of the self energy and compute its transverse 
traceless part. By writing a spectral representation for the self-energy,
and by using appropriately subtracted dispersion relations, we will be able to
compute the spectrum of the spin-2 bound states of the CFT. Upon coupling to
gravity, these bound states are analogous to the tower of massive Kaluza-Klein
gravitons that is found in the Karch-Randall compactification~\cite{kr}. 
By our computation of the self energy, we will explicitly check that even our 
free CFTs shares an essential feature with the KR model, namely, that the 
mass of these spin-2 states is $O(\sqrt{\Lambda})$. 
They are, therefore, heavier
than the graviton that undergoes a Higgs phenomenon, as the mass of the
latter is $\sim \sqrt{G}\Lambda \ll \sqrt{\Lambda}$.

Our analysis of a free CFT sheds light on the surprisingly many aspects that
it shares with strongly interacting CFTs, that can only be studied by 
resorting to holographic duality. These similarities suggest that 
many properties of gravity on an AdS background are due to the geometry of
the background, but are insensitive to many of the details of the matter 
CFT.

Our paper is organized as follows. In Section 2 we review the techniques of
Ref.~\cite{p3}, and we apply them to the computation of the 
transverse-traceless part of the 
self energy. In Section 3 we give the spectral representation of the self 
energy. This allows us to find the spectrum of spin-2 bound states formed by 
the CFT. We then write a (subtracted) 
dispersion relation for the self energy, and we use it to derive an alternative
expression for the self energy, 
that manifestly shows the high-energy behavior of the graviton propagator.
In Section 4 we present some concluding comments. 

\section{The Graviton Self-Energy} 
In this Section we briefly review the main ideas and techniques of~\cite{p3}
relevant for the computation of the graviton self-energy.

Consider a CFT consisting of a single free scalar coupled to Einstein gravity
 on $AdS_4$. Neglecting graviton loops, the two-point function of the 
stress-energy tensor coincides with the one-loop graviton self-energy 
$\Sigma$, 
\beq
 \Sigma^{\mu\nu,\rho\sigma}(x,y)\equiv\left.{\delta^2 \hat{W}[g]\over 
\delta g_{\mu\nu}(x) \delta g_{\rho\sigma}(y)}\right|_{g=\bar{g}},
\eeq{a1}
where the generating functional $\hat{W}[g]$ 
of the correlators of the stress-energy tensor is normalized by the condition
$\delta \hat{W}[g]/\delta g_{\mu\nu} |_{g=\bar{g}}=0$. 
An overbar is used hereafter to denote the background $AdS_4$ values.
Instead of $\Sigma$, it is more convenient to expand 
$g_{\mu\nu}=\bar{g}_{\mu\nu}+ h_{\mu\nu}$, and to compute the quadratic term in
the expansion of $\hat{W}[g]$ in powers of $h_{\mu\nu}$
\bea
\hat{W}[\bar{g}+h]&=&\hat{W}[\bar{g}]+ {1\over 2} h*\Sigma*h + O(h^3), 
\nonumber\\ 
h*\Sigma*h &=& \int d^4 x \sqrt{\bar{g}(x)}\int d^4 y \sqrt{\bar{g}(y)}
h^{\mu\nu}(x)\langle T_{\mu\nu}(x) T_{\rho\sigma}(y)\rangle h^{\rho\sigma}(y).
\eea{a2}
The stress-energy tensor of a free conformal scalar on  $AdS_4$ is given by
\beq
T_{\mu\nu}= \partial_\mu \phi \partial_\nu \phi -
{1\over 2}\bar{g}_{\mu\nu}\partial_\lambda\phi D^\lambda\phi - 
{1\over 6} [D_\mu D_\nu - \bar{g}_{\mu\nu}(\Box-\Lambda)]\phi^2.
\eeq{a3} 
On traceless, divergenceless tensors obeying 
$h^\mu_\mu=0$, $D_\mu h^{\mu\nu} =0$, Eq.~(\ref{a2}) reads 
\beq
h*\Sigma*h= \int d^4 x \sqrt{\bar{g}(x)}\int d^4 y \sqrt{\bar{g}(y)}
h^{\mu\nu}(x)\langle \partial_\mu \phi(x) \partial_\nu\phi (x)
\partial_\rho \phi (y)\partial_\sigma\phi(y) \rangle h^{\rho\sigma}(y).
\eeq{a4}
Using Wick's theorem, we can write Eq.~(\ref{a4}) as
\beq
h*\Sigma*h= 2\int d^4 x \sqrt{\bar{g}(x)}\int d^4 y \sqrt{\bar{g}(y)}
h^{\mu\nu}(x) h^{\rho\sigma}(y) {\partial\over \partial x^\mu} 
{\partial\over \partial y^\rho} \Delta (x,y)
 {\partial\over \partial x^\nu} 
{\partial\over \partial y^\sigma} \Delta (x,y) ,
\eeq{a5}
where  $\Delta (x,y)$ is a scalar propagator.

Further calculations are greatly facilitated by the use of the embedding
of $AdS_4$ in $R^5$~\cite{f} and by writing the integrals in Eq.~(\ref{a4}) in
five dimensions. 

In $R^5$ with the metric 
 $\eta=diag(-1,-1,+1,+1,+1)$ Anti de Sitter space is realized as the 
covering space of the hyperboloid $X^M X^N \eta_{MN}=-L^2$, $M,N=0,..,4$.
All  $SO(2,3)$-invariant quantities are functions of 
 $Z = X^M Y_M/L^2$ only. When standard (i.e. reflecting) 
boundary conditions are given, the scalar propagator is~\cite{aj} 
\beq
\Delta_E(Z)= {1\over 4\pi^2 L^2}{\Gamma(E)\Gamma(E-1)\over \Gamma(2E-2)}
 Z^{-E}\ofo \left(E,E-1;2E-2;1/Z\right),
\eeq{a6}
where $E$ labels the $SO(2,3)$-irreducible spin 0 representations
 $D(E,0)$~\cite{n}.
The 4-d tensor field  $h_{\mu\nu}$ can be promoted to a 5-d one by the 
usual embedding formula
\beq
h^{MN}(x)=\partial_\mu X^M(x) \partial_\nu X^N(x) h^{\mu\nu}(x).
\eeq{a7}
The important property $X^M h_{MN}(X)=0$ follows from the definition of 
$h_{MN}$.
A spin-2 field belongs to the representation $D(E,s)$. Its mass 
is related to its AdS energy $E$ by $\mu^2=L^{-2}E(E-3)$. It obeys the wave 
equation~\cite{f}
\beq
[X^2\partial^2 -(X\cdot \partial)^2 -3X\cdot \partial + E(E-3)]h_{MN}=0.
\eeq{m1}

The self-energy Eq.~(\ref{a4}) can now be written as a 5-d integral
\bea
h*\Sigma*h&=&2\int d\mu(X)\int d\mu(Y) h^{AB}(X)h^{CD}(Y) [\Delta'(Z)
 \eta^{AC}
+Y^AX^C\Delta''(Z)]\nonumber \\ &\,& [\Delta'(Z) \eta^{BD}
 +Y^BX^D\Delta''(Z)] ,
\eea{a8}
where $d \mu (X) \equiv \delta (X^2 + L^2) d^5 X$, and the prime denotes
 the derivative with respect to $Z$. In the case of a conformally coupled 
scalar, $\Delta$ is a linear superposition 
of $\Delta_1$ and $\Delta_2$, as explained later [see Eq.~(\ref{a13})]. 
From now on we set $L=1$.
Using the definition of $Z$ and integrating by parts, we reduce 
Eq.~(\ref{a8}) to
\beq
h*\Sigma*h=2\int d\mu(X)\int d\mu(Y) h^{AB}(X)h^{CD}(Y) Y_A X_C
{\partial\over \partial X^B}\Delta'(Z){\partial\over \partial Y^D}
\Delta'(Z).
\eeq{a9}
Introducing now the functions $G(Z)$ and $H(Z)$ by 
\beq
{\partial\over \partial X^A}\Delta'(Z) \Delta''(Z)
 \equiv Y_A G'(Z)\,,  \qquad
{\partial Z\over \partial X^A} G(Z) \equiv
 {\partial H(Z)\over \partial X^A}\,,
\eeq{a10}
we can write the self-energy in the following simple form
\beq
h*\Sigma*h=4\int d\mu(X)\int d\mu(Y) h^{AB}(X)H(Z)h_{AB}(Y)\,.
\eeq{a11}
The function $H(Z)$ obeys 
\beq
H''(Z) = (\Delta'')^2 .
\eeq{a12}
For a generic boundary condition on $AdS_4$, a conformal scalar belongs
to the irreducible representation $D(1,0)\oplus D(2,0)$~\cite{p3}.
Accordingly, the scalar propagator Eq.~(\ref{a6}) is a linear combination
\beq
\Delta (Z)= {1\over 4\pi^2}
\left(\alpha {1\over Z^2-1} + \beta {Z\over Z^2-1}\right)\,,
\eeq{a13}
where $\alpha=0, \beta=1$ ( $\alpha=1, \beta=0$) correspond to the scalar
being in  $D(1,0)$ [$D(2,0)$], respectively; this choice of $\alpha$, $\beta$
implies reflecting boundary conditions on $AdS_4$~\cite{ais,bf}.
The holographic interpretation of the KR model requires instead non-reflecting
boundary conditions, as emphasized in Refs.~\cite{p2,br}. They correspond to 
$\alpha = \beta = 1/2$~\cite{p3}.

The solution of Eq.~(\ref{a12})
can be written as
\beq
H_{(\alpha,\beta)}(Z) = (\alpha^2 +\beta^2) A(Z) + 2\alpha\beta B(Z) + 
 (\alpha^2 -\beta^2)\left[ C(Z) +D(Z)\right]\,,
\eeq{a14}
where the functions $A$, $B$, $C$, $D$ can be 
expressed in terms of the propagators
$\Delta_{1,2}$, $\Delta_{4}$, and their derivatives
\begin{equation}
A(Z) = {1\over 320\pi^4} \left[ {1\over (Z-1)^4} + {1\over (Z+1)^4} \right] =
 {1\over 10}\left[ (\Delta_1 ')^2 + (\Delta_2 ')^2  \right]\,,
\label{a15}
\end{equation}
\begin{equation}
B(Z) = {1\over 320\pi^4} \left[ {1\over (Z-1)^4} - {1\over (Z+1)^4} \right] =
 {\Delta_1 '\Delta_2 '\over   5}\,,
\label{a16}
\end{equation}
\begin{equation}
C(Z) = {1\over 64\pi^4 (Z^2-1)} = {1\over 16\pi^2} \Delta_2 (Z)\,,
\label{a17}
\end{equation}
\begin{equation}
D(Z)  = {3\over 128\pi^4}\left( Z \log{ {Z-1\over Z+1}} +2\right)
 = {1\over 16\pi^2} \left[ \Delta_4 (Z) - \Delta_2 (Z) \right]\,.
\label{a18}
\end{equation}
Combining Eqs.~(\ref{a15})-(\ref{a18}) we obtain
\begin{equation}
H_{(\alpha,\beta)}(Z) =  (\alpha^2 +\beta^2) { (\Delta_1 ')^2 +
 (\Delta_2 ')^2  \over 10} +
2\alpha\beta {\Delta_1 '\Delta_2 '\over   5}
 + (\alpha^2 -\beta^2) {1\over 16\pi^2} \Delta_4 (Z)\,.
\label{a19}
\end{equation}
The integration constants in Eq.~(\ref{a14}) were 
chosen so as 
to ensure $H_{(\alpha,\beta)}(Z)\rightarrow 0$ for $Z\rightarrow\infty$.
For $\alpha = \beta$ (in particular, for the KR model) we have simply
\begin{equation}
H_{(\alpha,\alpha)}(Z) = { \alpha^2 \left(\Delta_1 ' + 
 \Delta_2 '\right)^2 \over 5}\,.
\label{a20}
\end{equation}
In the next Section we shall obtain the spectral representation
 for Eq.~(\ref{a19}) and compute the self-energy.

\section{Spectral Representation for the Self-Energy}

The following spectral representation for  $H_{(\alpha,\beta)}(Z)$ can
be obtained by direct computation:
\begin{equation}
H_{(\alpha,\beta)}(Z)
 = \int\limits_{4}^{\infty} \rho_{(\alpha,\beta)} (E) \Delta (Z) dE\,,
\label{a21}
\end{equation}
where the spectral density is given by
\begin{eqnarray}
\rho_{(\alpha,\beta)} (E) &=& {\alpha^2 +\beta^2\over 480\pi^2}
 \sum_{n=0}^{\infty} 
(n+1)(n+2)(2n+1)(2n+3)(4n+5) \delta (E_0 -4-2n) \nonumber \\
&+& 
 {\alpha\beta\over 240\pi^2}
 \sum_{n=0}^{\infty} 
(n+1)(n+2)(2n+3)(2n+5)(4n+7) \delta (E_0 -5-2n) \nonumber \\
&+&  {\alpha^2 -\beta^2\over 16\pi^2}\delta (E_0 -4)\,.
\label{spectral_density}
\end{eqnarray}
Combining Eq.~(\ref{a11}) with Eqs.~(\ref{a21},\ref{spectral_density}) 
we obtain the graviton self-energy as follows. Call $h_{MN}^{(\mu)}$ the 
transverse traceless spin-2 field of mass $\mu$, obeying Eq.~(\ref{m1}). 
Define the function  
$\tilde{\Sigma}_{(\alpha,\beta)}(\mu^2)$ by
\beq
h^{(\mu)} *\Sigma_{(\alpha,\beta)}*h^{(\mu)}
=\int d\mu(X) h^{(\mu)}_{AB}(X)h^{(\mu)\;AB}(X)
\tilde{\Sigma}_{(\alpha,\beta)}(\mu^2).
\eeq{m2}
The transverse-traceless part of the self-energy on states of mass $\mu$ is
equal to $\tilde{\Sigma}_{(\alpha,\beta)}(\mu^2)$, which reads, explicitly 
\begin{eqnarray}
\tilde{\Sigma}_{(\alpha,\beta)}(\mu^2) &=& {\alpha^2 +\beta^2\over 120\pi^2}
 \sum_{n=0}^{\infty} 
{(n+1)(n+2)(2n+1)(2n+3)(4n+5)\over (2n+4)(2n+1) -\mu^2}  \nonumber \\
&& +
 {\alpha\beta\over 60\pi^2}
 \sum_{n=0}^{\infty} 
{(n+1)(n+2)(2n+3)(2n+5)(4n+7)\over (2n+5)(2n+2) - \mu^2} \nonumber \\
&& + { \alpha^2 -\beta^2 \over 4\pi^2 (4 - \mu^2)}.
\label{a22}
\end{eqnarray}
From this equation, it is clear that the choice of boundary conditions on 
$AdS_4$ influences the spin 2 bound-state spectrum of the theory.
For generic $\alpha$, $\beta$ the spectrum consists of two series,
\beq
\mu^2_k L^2 = (2k+1)(2k+4) = 4, 18, 40, ...\,,
\eeq{m3}
\beq
\mu^2_k L^2 = (2k+2)(2k+5) = 10, 28, 54, ...\,. 
\eeq{m4}
In these formulas, we reintroduced the AdS radius $L$ for clarity. Notice that
the second series of poles, i.e. of spin-2 bound states of the CFT, only
exists for non-reflecting boundary conditions.

Introducing now $t=\mu^2$, $t'=m^2$, we can write (\ref{a22})
as a contour integral.
Define first
\begin{equation}
s_{\pm}(t') = {3\over 2 } \left( 1\pm \sqrt{1+ {4t'\over 9}}\right)\,.
\label{a24}
\end{equation}
Then Eq.~(\ref{a22}) can be written as
\begin{equation}
\tilde{\Sigma}_{(\alpha,\beta)} (t) = 
{1\over 2\pi i} \int_C {{\cal M}_{(\alpha,\beta)} (t')dt'
\over t' - t}\,,
\label{a25}
\end{equation}
where 
\begin{eqnarray}
{\cal M}_{(\alpha,\beta)} (t)  &=& - {\alpha^2 +\beta^2
 \over 2\pi i \, 960 \pi^2}
 \left[ \psi \left( {4-s_+(t)\over 2}\right) +
 \psi\left( {4-s_-(t)\over 2}\right)\right]
 t (t+2) \nonumber \\
&&-
{ \alpha\beta
 \over 2\pi i \, 480 \pi^2}
 \left[ \psi \left( {5- s_+(t)\over 2}\right) +
 \psi\left( {5-s_-(t)\over 2}\right)\right] t (t+2) \nonumber \\
&&+
{\alpha^2 -\beta^2 \over 4 \pi^2 (4 - t)}\,,
\label{a26}
\end{eqnarray}
and the closed contour $C$ includes all and only the poles given in
Eqs.~(\ref{m3},\ref{m4}).

The representation given in Eq.~(\ref{a25}) is the first step toward a 
Sommerfeld-Watson resummation. The second step would be to deform the contour
of integration to another contour, $C'$, that only includes the pole at 
$t'=t$. The problem encountered in attempting to resum Eq.~(\ref{a22}) is 
that the dispersion integral Eq.~(\ref{a25}) needs subtractions, since 
${\cal M}_{(\alpha,\beta)} (t)$ does not vanish for $|t|\rightarrow \infty$.
Following a standard method of field theory (see e.g.~\cite{landau})
we write a dispersion integral using 
${\cal M}_{(\alpha,\beta)} (t)/t^3$. 
We get
\beq
\tilde{\Sigma}_{(\alpha,\beta)} (t) = A + Bt + Ct^2 +
{t^3\over \pi}  \int_C 
{{\cal M}_{(\alpha,\beta)} (t')
\over t'^3 (t' - t )}dt'\,.
\eeq{m5}
The constants $A,B,C$ are arbitrary since they arise from local (contact) 
terms in $\hat{W}[g]$. In particular, $B$ is proportional to the coefficient
multiplying the Einstein action $\int d^4x \sqrt{g} (R-2\Lambda)$. In units 
$L=1$ the cosmological constant is $\Lambda=-3$. $C$ is proportional to the 
coefficient multiplying the square of the Weyl tensor, 
$\int d^4 x \sqrt{g} C_{\mu\nu\rho\sigma}C^{\mu\nu\rho\sigma}$. 
Finally, $A$ is proportional to the graviton mass.
Since $A$, $B$, $C$ are arbitrary, we need additional 
information to determine them. This is why, in particular, our computation 
cannot tell by itself whether the graviton gets a mass in $AdS_4$. 
To find if $A$ is nonzero, one must study instead the longitudinal parts of 
the graviton propagator,
and find out if they have a pole corresponding to a spin-1 Goldstone 
boson~\cite{p3}~\footnote{In $AdS_4$ that Goldstone boson is massive because
it belongs to the $D(4,1)$ representation of $SO(2,3)$~\cite{p3}.}.
  
Now the integrand ${\cal M}_{(\alpha,\beta)} (t')/ t'^3$ vanishes at 
$|t'|=\infty$ so that the contour of integration
$C$ can be deformed without picking up unwanted terms at infinity. 

By explicit computation of the integral in Eq.~(\ref{m5}) we get 
\begin{eqnarray}
\tilde{\Sigma}_{(\alpha,\beta)}(\mu^2) &=& 
 - { \mu^2 (\mu^2+2)\over 960 \pi^2}
\Biggl\{   (\alpha^2 +\beta^2) \left[ 
\psi \left( {5\over 4} + {\sqrt{9+4\mu^2}\over 4}\right) +
\psi \left( {5\over 4} - {\sqrt{9+4\mu^2}\over 4}\right)\right]\nonumber \\
&& +2\alpha\beta  \left[ 
\psi \left( {7\over 4} + {\sqrt{9+4\mu^2}\over 4}\right) +
\psi \left( {7\over 4} - {\sqrt{9+4\mu^2}\over 4}\right) 
\right]\Biggr\}\nonumber \\
&& +{a^2 \mu^6  (\alpha^2 -\beta^2)\over 256 \pi^2 (4-\mu^2)} + P(\mu^2)\,,
\end{eqnarray}
where $P(\mu^2)$ is an rather cumbersome polynomial in $\mu^2$ given by
\begin{eqnarray}
 P(\mu^2) &=& A + B\mu^2 + C\mu^4 +
{\mu^2 \alpha\beta [ 54(\mu^2+2)\psi(5/2) +\mu^2 (9\pi^2-80)]
\over 25920\pi^2} \nonumber \\ &&-
{\mu^2 (\alpha^2 +\beta^2)[\mu^2 \pi^2 + 6(\mu^2+2)(\gamma_E +2\log{2})]
\over 5760\pi^2}\nonumber \\
&&+
 {\mu^2  (\alpha^2 +\beta^2) [36 - 18\gamma_E (\mu^2+2) +\mu^2 (12+\pi^2)]
\over 17280\pi^2}\nonumber \\
&&-
 {\mu^2 \alpha\beta [18\gamma_E (\mu^2 + 2) + \pi^2\mu^2]\over 8640\pi^2}\,.
\end{eqnarray}

For small $\mu$ we have
\bea
\tilde{\Sigma}_{(\alpha,\beta)} &=& A +B\mu^2 + C\mu^4 +
{\mu^6\over 3732480\pi^2}
\Biggl[ 3933\alpha^2 + 896 \alpha\beta - 3357\beta^2 \nonumber \\ && +
 168\pi^2 (\alpha - \beta )^2 +1728\zeta (3)
 (\alpha +\beta)^2\Biggr] + O \left( \mu^8 \right)\,.
\eea{m6}
Notice that the first unambiguous term here is $O(\mu^6)$, and that there is 
no logarithmic term in $\mu$. This is not surprising since in Eq.~(\ref{m6})
we have expanded our formulas for $\mu \ll 1$. The flat-space form of the
self-energy, proportional to $\mu^4\log \mu^2$ arises in the opposite limit
$\mu \gg 1$. Finally, to reinstate $L$ in all the previous formulas 
one must perform the substitution $\mu \rightarrow L\mu$.

\section{Conclusions}
In this letter, we coupled Einstein gravity on $AdS_4$ 
to a very simple CFT, made of a single free, conformally coupled scalar.
Even in this case we found an interesting phenomenon, that has no counterpart
in Minkowski space, namely, the appearance of a new tower of spin-2 bound
states any time the scalar field is given non-reflecting boundary conditions.
This ``bulk'' effect of conditions given at the boundary of $AdS_4$ paralles
the finding in~\cite{p3} and has no flat-space counterpart. In~\cite{p3},
it was found that the massless graviton of AdS Einstein gravity acquires
a mass when gravity is coupled to conformal matter, {\em and conformal matter
is given non-reflecting boundary conditions}. It may seem surprising that, by 
doing a local experiment, one can determine what happens at the boundary of
space. The surprise is lessened by noticing that in AdS 
light takes only a finite coordinate time to travel from a point inside the 
space to the boundary. Also, the precision in mass measurements required to
tell boundary conditions from the spin-2 spectra in Eqs.~(\ref{m2},\ref{m3}) is
$O(1/L)$. This means that any such measurement has to be performed either over
horizon scales or over a Hubble time. 
\vskip .2in
{\bf Acknowledgments}\vskip .1in
\noindent
We would like to thank N.N. Khuri and C. Fronsdal for interesting discussions. 
This work is supported in part by NSF grant PHY-0070787.

\end{document}